\documentstyle[12pt]{article}
\begin{document}
\thispagestyle{empty}
\begin{raggedleft}
IF-UFRJ-nnn/96\\
hep-th/9608149\\
July/96\\
\end{raggedleft}
$\phantom{x}$\vskip 0.618cm\par
\begin{center}{\Huge Soldering Chiralities II:
Non-Abelian Case}
\vspace{2cm}
%\vfill
%\begin{center}
$\phantom{X}$\\
{\Large Clovis Wotzasek}\footnote{EMAIL:CLOVIS@IF.UFRJ.BR}\\[3ex]
{\em Instituto de F\'\i sica\\
Universidade Federal do Rio de Janeiro\\
21945, Rio de Janeiro, Brazil\\}
\end{center}\par
%\vfill
\vspace{2cm} 
%\maketitle
\abstract

\noindent 
We study the non-abelian extension of the soldering process of two chiral WZW models of opposite chiralities, resulting in a (non-chiral) WZW model living in a 2D space-time with non trivial Riemanian curvature.

%\vspace{2cm}
%\noindent PACS: 

\newpage
%\section{Introduction}
\noindent{\bf 1}   In a recent work, Stone\cite{stone} has shown that the method of coadjoint orbit\cite{kirilov}, when applied to a representation of a group associated with a single affine Kac-Moody algebra, provides an action for the chiral WZW model, or Sonneschein chiral boson\cite{sonneschein}, which is a non-abelian generalization of the Floreanini and Jackiw model for chiral scalar fields\cite{florjack}.  This method provides an useful bosonization scheme for Weyl fermions, since a level one representation of LU(N) has an interpretation as the Hilbert space for a free chiral fermion\cite{segal}.

The drawback of this approach is that only Weyl fermions can be dealt with in this way, since 2D conformally invariant QFT has separate right and left current algebras.  This is in contrast to the result of Gepner and Witten\cite{gepner}, showing that necessity for modular invariance leads to the same representation for the right and left affine KM algebras in the WZW model\cite{witten}.

In order to overcome this difficulty, Stone\cite{stone2} proposed to solder two chiral models, introducing a non-dynamical gauge field to remove the degree of freedom that obstructs the vector gauge invariance.  He observed that the equality for the weights in the two representations is physically connected with the need to abandon one of the two separate chiral symmetries, and accept that only vector gauge symmetry should be mantained.

More recently, this author and collaborators\cite{solda} showed that soldering two Siegel models\cite{siegel} of opposite chiralities results in an action for the scalar field immersed on a 2D space-time with non-trivial Riemanian curvature.  In this Note, we extend that result to non-abelian fields, and show that soldering, a la Stone, two chiral WZW models results in a WZW model coupled to gravity.  The result so obtained implies that the effective action that comes out of the soldering process is invariant under the full diffeomophism group, which is not a mere sum of two Siegel symmetries.\vspace{0.3cm}\\

%\section{Chiral decomposition for the abelian field}

\noindent{\bf 2}    To begin with, let us review some known facts about the non-abelian Siegel model\cite{frishson}.  The action for a left mover particle is given as\footnote{Our notation is as follows: $x^{\pm}={1\over 2}(t\pm x)$ are the light-cone variables and $\tilde g = g^{-1}$ is the inverse matrix.}

\begin{equation}
\label{leftzero}
S_0^{(+)}(g)=\int\;d^2x\; tr\left(\partial_+g\:\partial_-\tilde g+\lambda_{++}\partial_-g\:\partial_-\tilde g\right) +\Gamma_{WZ}(g)
\end{equation}

\noindent where $g\in G$ is a matrix-valued field taking values on some compact Lie group $G$, with an algebra ${\bf G}$.   The term $\Gamma_{WZ}(g)$ is the topological Wess-Zumino functional, as defined in Ref.\cite{polywieg}.  This action can be seen as the WZW action, immersed in a gravitational background, with an appropriately truncated metric tensor:

\begin{equation}
\label{leftzerograv}
S_0^{(+)}(g)={1\over 2}\int d^2 x \sqrt{-\eta_+}\; \eta_+^{\mu\nu}\:tr\left(\partial_\mu g\:\partial_\nu \tilde g\right) +\Gamma_{WZ}(g)
\end{equation}

\noindent with $\eta^+ = det(\eta^+_{\mu\nu})$ and

\begin{equation}
\label{metric+}
{1\over 2} \sqrt{-\eta_+}\: \eta_+^{\mu\nu}=\left(
\begin{array}{cc}
0 & {{1\over 2}}\\
{{1\over 2}} & {\lambda_{++}}
\end{array}
\right)
\end{equation}

\noindent  Out of the two affine invariances of the original WZW model, only one is left over due to the chiral constraint $\partial_- g\approx 0$.  To see this we compute the Noether currents for the axial, vectorial and chiral transformations.  The variation of the Siegel-WZW action (\ref{leftzerograv}) gives

\begin{equation}
\label{variation}
\delta S_0^{(+)}(g)=\left\{
\begin{array}{ll}
\int d^2x\:tr\left\{\delta g \tilde g\: 2\left[\partial_+\left(\partial_- g\tilde g\right)\:+\:\partial_-\left(\lambda_{++}\partial_- g\tilde g\right)\right]\right\}\\
\int d^2x\:tr\left\{\tilde g\delta g \: 2\left[\partial_-\left(\tilde g\partial_+ g\right)\:+\:\partial_-\left(\lambda_{++}\tilde g\partial_- g\right)\right]\right\}
\end{array}\right.
\end{equation}

\noindent  From (\ref{variation}) and the axial transformation ($g\rightarrow kgk $), we obtain

\begin{eqnarray}
\label{axialcurr}
J_A^+&=& 2g\partial_-\tilde g\nonumber\\
J_A^-&=& -2\left[\tilde g\partial_+ g +\lambda_{++}(\tilde g\partial_- g +\partial_-g \;\tilde g)\right]
\end{eqnarray}

\noindent where $k \in H $ take their values in some subgroup $H$ of $G$.  For the vectorial transformation ($ g \rightarrow \tilde k g k$) we obtain

\begin{eqnarray}
\label{vectcurr}
J_V^+&=& 2g\partial_-\tilde g\nonumber\\
J_V^-&=& 2\left[\tilde g\partial_+ g +\lambda_{++}(\tilde g\partial_- g -\partial_-g \;\tilde g)\right]
\end{eqnarray}

\noindent One observes that the axial and the vectorial currents above are dual to each other in the extended sense

\begin{equation}
\mbox{}^*T^\mu=\sqrt{-\eta} \; \eta^{\mu\nu}\epsilon_{\mu\lambda}T^\lambda
\end{equation}

\noindent as can be easily checked.  Also, using the truncate metric $\eta_+^{\mu\nu}$ to lower spacetime indices, we get

\begin{eqnarray}
J_+ &=& J^- -2 \lambda_{++}J^+\nonumber\\
J_- &=& J^+
\end{eqnarray}

\noindent  The chiral currents can be obtained either from the left ($g\rightarrow gk$) and
right ($g\rightarrow \tilde k g$) transformation, or from the definition

\begin{eqnarray}
J_L^\mu={1\over 2}\left(J_A^\mu - J_V^\mu\right)\nonumber\\
J_R^\mu={1\over 2}\left(J_A^\mu + J_V^\mu\right)
\end{eqnarray}

\noindent  The result is 

\begin{eqnarray}
J_L^{(+)} &=& 0\nonumber\\
J_L^{(-)} &=& 2\left(\tilde g \partial g +\lambda_{++} \tilde g\partial_- g\right)
\end{eqnarray}

\noindent and

\begin{eqnarray}
J_R^{(+)} &=& -2 g\partial_-\tilde g\nonumber\\
J_R^{(-)} &=& -2\lambda_{++} g \partial_-\tilde g
\end{eqnarray}

\noindent  As mentioned, the affine invariance is only present in the left sector since $J_L^{(+)}=0$ and $\partial_-J_L^{(-)}=0$, which implies $J_L^{(-)}=J_L^{(-)}(x^+)$, while $J_R^{(-)}\neq 0$ and $J_R^{(+)}\neq J_R^{(+)}(x^-)$.

Similarly, for the other chirality, we have

\begin{eqnarray}
S_0^{(-)}(h)&=&\int \;d^2x\; tr\left(\partial_+h\partial_-\tilde h +\lambda_{--}\partial_+ h\partial_+\tilde h\right) - \Gamma_{WZ}(h)\nonumber\\
&=& {1\over 2} \int d^2x\; \sqrt{-\eta_-} \;\eta_-^{\mu\nu}\:tr\left(\partial_\mu h \partial_\nu 
\tilde h\right)- \Gamma_{WZ}(h)
\end{eqnarray}

\noindent  where

\begin{equation}
{1\over 2}\sqrt{-\eta_-}\:\eta_-^{\mu\nu}=\left(
\begin{array}{cc}
{\lambda_{--}} & {{1\over 2}}\\
{{1\over 2}} & 0
\end{array}
\right)
\end{equation}

\noindent  The set of axial, vector and chiral Noether currents are

\begin{eqnarray}
J_A^{(+)}(h)&=& 2\left[\tilde h\partial_- h +\lambda_{--}\left(\tilde h\partial_+ h +\partial_+ h \tilde h\right)\right]\nonumber\\
J_A^{(-)}(h)&=&2\:\partial_+ g\tilde g\\
J_V^{(+)}(h)&=&2\left[\tilde h\partial_- h +\lambda_{--}\left(\tilde h\partial_+ h -\partial_+ h\:\tilde h\right)\right]\nonumber\\
J_V^{(-)}(h)&=&-2\:\partial_+ h\:\tilde h\\
J_L^{(+)}(h)&=& 2\:\partial_+ h\:\tilde h\nonumber\\
J_L^{(-)}(h)&=&2\:\lambda_{--}\:\partial_+ h\:\tilde h\\
J_R^{(+)}(h)&=&2\left(\tilde h\partial_- h +\lambda_{--}\:\tilde h\partial_+ h\right)\nonumber\\
J_R^{(-)}(h)&=&0
\end{eqnarray}

Next, let us consider the gauging procedure to be adopted in the soldering of the right and the left Siegel actions.  As discussed in \cite{gauging}, the coupling with gauge fields is only consistent if used the correspondent chiral current, otherwise the equations of motion, after the gauging, will result being incompatible with the covariant chiral constraint.  This can be done quite simply by using an iterative Noether procedure to gauge the global (left) chiral transformation

\begin{eqnarray}
\label{transf.1}
g&\rightarrow & gk\nonumber\\
\lambda_{++}&\rightarrow & \lambda_{++}\nonumber\\
A_-&\rightarrow & \tilde k A_-k + \tilde k\partial_- k
\end{eqnarray}

\noindent  To compensate for the non-invariance of $S_0^{(+)}$, we introduce the coupling term

\begin{equation}
S_0^{(+)}\rightarrow S_1^{(+)} =S_0^{(+)} +A_- J_L^-(g)
\end{equation}

\noindent  along with the gauge field $A_- $, taking values in the subalgebra ${\bf H}$ of $H$, whose transformation properties are being defined in (\ref{transf.1}).  Using the transformations laws above, it is a simple algebra to find

\begin{equation}
\delta\left(S_1^{(+)}-\lambda_{++} A_-^2\right)=2\;\partial_+\omega\; A_-
\end{equation}

\noindent where $\omega \in {\bf H}$ is an infinitesimal element of the algebra ${\bf H}$.  One can see that $S_2^{(+)}=S_1^{(+)} -\lambda_{++}\;A_-^2$ cannot be made gauge invariant by addtional Noether counter-terms, but it has the virtue of being independent on the transformations properties of $g$.  Similarly, for the opposite chirality we find

\begin{equation}
\delta S_2^{(-)}=-\;2\;A_+\;\partial_-\omega
\end{equation}

\noindent for

\begin{equation}
S_2^{(-)}(h)=S_0^{(-)}(h)-A_+J_R^{+}(h)-\;\lambda_{--}\;A_+^2
\end{equation}

\noindent  when the fields transform as

\begin{eqnarray}
\label{transf.2}
h&\rightarrow &hk\nonumber\\
A_+&\rightarrow & kA_+\tilde k + k\partial_+\;\tilde k\nonumber\\
\lambda_{--} &\rightarrow & \lambda_{--}
\end{eqnarray}

\noindent  Altough the gauged actions for each chirality cannot be made gauge invariant separately, one can show that, with the inclusion of a contact term, their sum can.  Indeed, the combined action

\begin{equation}
\label{eff}
S_{tot}=S_2^{(+)} + S_2^{(-)} + 2 A_+ \; A_-
\end{equation}

\noindent is invariant under the set of transformations (\ref{transf.1}) and (\ref{transf.2}).

Following Ref.\cite{stone}, we eliminate the (non dynamical) gauge field $ A_\mu $.  From the equations of motion one gets

\begin{equation}
{\cal J} = 2 {\bf M} {\cal A}
\end{equation} 

\noindent with

\begin{equation}
{\cal J} =\left(
\begin{array}{c}
J_L^-(g)\\
J_R^+(h)
\end{array}
\right)
\end{equation}

\begin{equation}
{\cal A} =\left(
\begin{array}{c}
{A_+}\\
{A_-}
\end{array}
\right)
\end{equation}

\noindent and

\begin{equation}
{\bf M} =\left(
\begin{array}{cc}
1 & {\lambda_{++}}\\
{\lambda_{--}} & 1
\end{array}
\right)
\end{equation}

\noindent  Bringing these results into the effective action (\ref{eff}) gives,

\begin{eqnarray}
S_{eff} &=& S_0^{(+)}(g) + S_0^{(-)}(h) +\nonumber\\
&+& \int d^2x {1\over{1-\lambda^2}} tr\left\{ 2\left[\tilde g\partial_+ g \;\tilde h\partial_- h + \lambda^2 \;\tilde g\partial_- g\;\tilde h\partial_+ h +\right.\right.\nonumber\\
&+& \left. \lambda_{++}\;\tilde g\partial_- g \;\tilde h\partial_- h +\lambda_{--}\;\tilde g\partial_+ g\;\tilde h\partial_- h \right] +\nonumber\\
&+& \lambda_{--}\;\left(\partial_+ g\;\partial_+\tilde g + 2\lambda_{++}\;\partial_+ g\;\partial_-\tilde g +\lambda_{++}^2\;\partial_- g\;\partial_-\tilde g\right) +\nonumber\\
 &+& \left. \lambda_{++}\;\left(\partial_- h\;\partial_-\tilde h + 2 \lambda_{--}\;\partial_+ h\;\partial_+\tilde h +\lambda_{--}^2\;\partial_+ h\;\partial_+\tilde h\right)\right\}  
\end{eqnarray}

\noindent where $\lambda^2=\lambda_{++}\lambda_{--} $.  To finish, we show that this effective action correspond to that of a (non-chiral) WZW model.  To this end we define a new metric tensor as,

\begin{equation}
{1\over 2}\sqrt{-\eta}\; \eta^{\mu\nu}= {1\over{1-\lambda^2}}\;\left(
\begin{array}{cc}
{\lambda_{--}} & {{{1+\lambda^2}\over 2}}\\
{{{1+\lambda^2}\over 2}} & {\lambda_{++}}
\end{array}
\right)
\end{equation}

\noindent and a new (effective) field

\begin{equation}
{\cal G}= g\tilde h
\end{equation}

\noindent Now, making use of the property of the Wess-Zumino functional $\Gamma_{WZ}(h)\:=\:-\;\Gamma_{WZ}(\tilde h)$, and of the Polyakov-Wiegman identity\cite{polywieg},it becomes a simple algebra to show that the effective action above can be rewritten as

\begin{equation}
S={1\over 2}\int \:d^2x\:\sqrt{-\eta}\:\eta^{\mu\nu}\:tr\left(\partial_\mu {\cal G}\;\partial_\nu\tilde{\cal G}\right) +\Gamma_{WZ}({\cal G})
\end{equation}

\noindent which shows that the effective field ${\cal G}$ satisfy a WZW model coupled minimally to gravity, which is the result claimed.\vspace{0.3cm}\\

\noindent ACKNOWLEDGEMENTS.  The author would like to thank the hospitality of the Department of Physics and Astronomy of University of Rochester, where part of this work was done.  The author is partially suported by CNPq, FINEP and FUJB , Brasil.


\begin{thebibliography}{30}

\bibitem{stone}M.Stone, Phys.Rev.Lett. 63 (1989) 731; Nucl.Phys. B327 (1989) 399.
\bibitem{kirilov}A.Kirilov, "Elements of the Theory of Representations", Springer-Verlag, Berlin, 1976.
\bibitem{sonneschein}J.Sonnenschein, Nucl.Phys. B309 (1988) 752.
\bibitem{florjack}R.Floreanini and R.Jackiw, Phys.Rev.Lett. 59 (1987) 1873.
\bibitem{segal}A.Pressley and G.Segal, "Loop Groups", Oxford Univ.Press, Oxford, 1986.
\bibitem{gepner}D.Gepner and E.Witten, Nucl.Phys. B278 (1986) 493.
\bibitem{witten}E.Witten, Comm.Math.Phys. 92 (1984) 455.
\bibitem{stone2}M.Stone, Illinois Report, ILL-TH-89-23.
\bibitem{solda}R.Amorim, A.Das, and C.Wotzasek, Phys.Rev. D53 (1996) 5810.
\bibitem{siegel}W.Siegel, Nucl.Phys. B238 (1984) 307.
\bibitem{frishson}Y.Frishman and J.Sonnenschein, Nucl.Phys. B301 (1988) 346.
\bibitem{polywieg}A.Polyakov and P.B.Wiegman, Phys.Lett B131 (1983) 121; Phys.Lett. B141 (1984) 223.
\bibitem{gauging}C.Wotzasek, "A remark on gauging chiral bosons", UFRJ Report/96.


\end{thebibliography}
\end{document}